# CHANNEL ESTIMATION STUDY FOR BLOCK - PILOT INSERTION IN OFDM SYSTEMS UNDER SLOWLY TIME VARYING CONDITIONS

Aida Zaier[1] and Ridha Bouallègue[2]

[1]National Engineering School of Tunis, Tunis University, Tunisia,
zaieraida@yahoo.fr
[2] High School of Communications, Tunis Ariana, Tunisia
ridha.bouallegue@gnet.tn

## ABSTRACT

*In this paper, we propose a study of performance of the channel estimation using LS, MMSE, LMMSE and Lr-LMMSE algorithms in OFDM (Orthogonal Frequency Division Multiplexing) system which, as known suffers from the time variation of the channel under high mobility conditions, using block pilot insertion.*

*The loss of sub channel orthogonality leads to inter-carrier interference (ICI). Using many algorithms for channel estimation, we will show that, for a 16- QAM modulation, the LMMSE algorithm performs well to achieve this estimation but when the SNR (Signal Noise Rate) is high, the four algorithms (LS, MMSE, LMMSE and Lr-LMMSE) perform similarly, this is not always the case for another scheme of modulation. We will improve also the mean squared error for these algorithms. It will be illustrious in this paper that the LMMSE algorithm performs well with the block- pilot insertion as well as its low rank version which behave very good even when the size of FFT is very high.*

## KEYWORDS

*OFDM Systems, Channel Estimation, block – pilot insertion, time varying channel, LS, MMSE, LMMSE, Lr-LMMSE.*

## 1. INTRODUCTION

Multicarrier modulations attract a lot of attention ranging from wireline to wireless communications compared to single carrier modulation because of theirs capability to efficiently cope with frequency selective fading channels. Much of attention in the present literature emphasizes on the use of conventional OFDM, which is able to avoid both inter - symbol interference (ISI) and inter - channel interference (ICI) making use of a suitable cyclic prefix.

The fundamental phenomenon which makes reliable wireless transmission difficult is the multipath fading. The main advantage of OFDM transmission is its robustness to frequency selective fading characteristics of a mobile radio channel. In OFDM, the entire signal bandwidth is divided into a number of narrow bands or orthogonal subcarriers, and signal is transmitted in the narrow bands in parallel. Therefore, it reduces inter symbol interference (ISI) and eliminates the need for complex equalization. [1]

OFDM is widely used in the wireless systems such as wireless LAN, terrestrial digital television broadcasting, cell-phone, and WiMAX.

In practice, a wideband radio channel is time-variant, frequency-selective and noisy. The estimation of its transfer function becomes rather difficult. First, the slow fading assumption does not always hold. Thus the transfer function might have significant changes even for adjacent OFDM data blocks. Therefore, it is preferable to estimate channel based on the pilots signals in each individual OFDM data block. Secondly, the pilot signals are also corrupted by inter-carrier interference (ICI), due to the fast variation of the mobile channel. In addition, additive white Gaussian noise (AWGN) always exists. IC1 and AWGN components in the received pilot signals strongly affect the accuracy of the estimation.

In general, some pilot signals are inserted as a reference for OFDM channel estimation and whole one OFDM frame is often used as a pilot frame. [2]

Focusing on the one dimensional estimation based on pilot insertion, we follow mainly block pilot insertion (which will be discussed in this paper), comb pilot insertion without forgetting those used on two dimensional estimation.

In this paper, we will improve mainly channel estimation for block type insertion based on LMMSE algorithm; we will compare also the channel estimators using the four algorithms LS, MMSE, LMMSE and Lr –LMMSE.

The main algorithm that will be enhanced in this paper is the linear minimum mean square error (LMMSE) for Orthogonal Frequency Division Multiplexing (OFDM) systems and its low rank version. The proposed channel estimation method requires the statistic knowledge of the channel in advance.

Generally, the current channel estimation methods can be classified into two categories. The first one is based on the pilots [3–4], and the second one is based on blind channel estimation [5–6] which does not use pilots and are not suitable for applications with fast varying fading channels. And most useful communication systems nowadays adopt pilot arrangement based channel estimation, for these reasons, this work studies the first category.

Channel estimation methods based on the pilot insertion can be divided into two classical pilot models, which are block-type and comb-type model [7]. The first model refers to that the pilots are inserted into all the subcarriers of one OFDM symbol with a certain period. The block-type can be adopted for slowly fading channel, that is, the channel can be considered as stationary within a certain period of OFDM symbols [7]. Nevertheless, the second model refers to that the pilots are positioned at some definite subcarriers in each OFDM symbol.

In this study, we have introduced also the estimation of the channel by the minimum mean-squared error which as known has a good performance but high complexity [8]. This estimator had demonstrated also a good behaviour in the case of the block - pilot insertion but still performs lower than the LMMSE algorithm.

Though a linear minimum mean-squared error (LMMSE) estimator using only frequency correlation has lower complexity than one using both time and frequency correlation, it still requires a large number of operations. We introduce a low-complexity approximation to a frequency-based LMMSE estimator that uses the theory of optimal rank reduction. [3][9]

This paper is organized as follows. Section II describes the system model. In section III, channel estimation scheme is analysed for the different algorithms proposed for this study. Finally simulation results are given in section IV and performed in terms of Bit Error Rate and Mean Square Error. Finally the conclusion is drawn in section V.

## 2. SYSTEM MODEL

The system model based on pilot channel estimation is depicted in Figure 1.

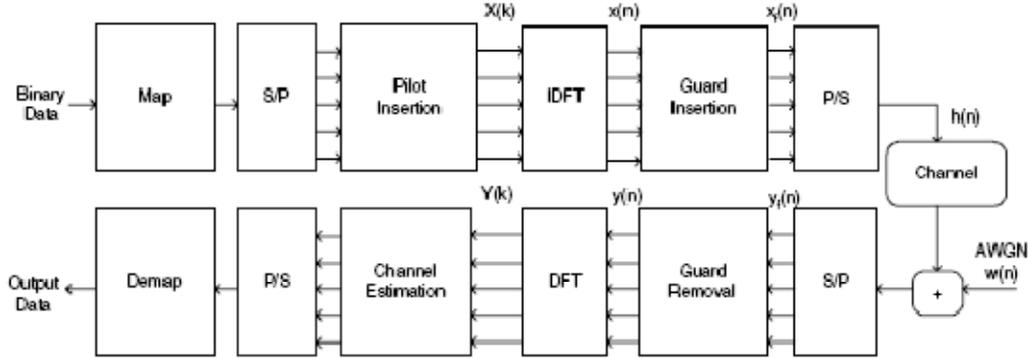

Figure 1: Baseband OFDM System [10]

The binary information is first grouped and mapped according to the modulation for signal mapping. Then, pilots will be inserted to all sub-carriers uniformly between the information data sequence or with a specific period. Yet, IDFT block is used for transforming the data sequence of length N{X(k)} into time domain signal {x(n)} as follow:

$$x(n) = IDFT\{X(k)\} \quad n = 0, 1, 2, ..., N-1$$
$$= \sum_{k=0}^{N-1} X(k) e^{j(2\pi kn/N)} \tag{1}$$

Where N is the DFT length. Subsequent the DFT block, the guard interval, which is chosen to be grater than the delay spread and contains the cyclically extended part of the OFDM symbol for eliminating inter-carrier interference, is inserted to avoid inter-symbol interference. The OFDM symbol resulting from this succession is the follow:

$$x_f(n) = \begin{cases} x(N+n), & n = -N_g, -N_g+1, ..., -1 \\ x(n) & n = 0, 1, ..., N-1 \end{cases} \tag{2}$$

Where $N_g$ is the length of the guard interval. Then, the OFDM symbol $x_f(n)$ will pass through the channel which is expected to be frequency selective and time varying with Rayleigh fading and an Additive White Gaussian Noise AWGN $w(n)$. The received signal is given by:

$$y_f(n) = x_f(n) \otimes h(n) + w(n) \tag{3}$$

Wherever $h(n)$ is the channel impulse response which can be represented as follow [11]:

$$h(n) = \sum_{i=0}^{r-1} h_i e^{j(2\pi/N) f_{Di} T_n} \delta(\lambda - \tau_i) \qquad 0 \le n \le N-1 \tag{4}$$

Where r is the total number of propagation paths, $h_i$ is the complex impulse response of the *i*th path, $f_{Di}$ is the *i*th path Doppler frequency shift, $\lambda$ is the delay spread index, T is the sample period and $\tau_i$ is the *i*th path delay normalized by the sampling time.

Then, at the receiver, after passing to discrete domain through S/P block, guard time is removed and the expression of y(n) is given by:

$$y_f(n) \text{ for } -N_g \le n \le N-1$$
$$y(n) = y_f(n+N_g) \quad n = 0, 1, 2, ..., N-1 \tag{5}$$

Then y(n) is driven to the DFT block and given by:

$$Y(k) = DFT\{y(n)\} \qquad k = 0, 1, 2, ..., N-1 \tag{6}$$
$$= \frac{1}{N} \sum_{n=0}^{N-1} y(n) e^{-j(2\pi kn/N)}$$

Within the framework of a transmission without inter-symbol interference ISI, the relation between the resulting Y(k) to $H(k) = DFT\{h(n)\}$, is given by[12]:

$$Y(k) = X(k)H(k) + I(k) + W(k) \tag{7}$$

Where I(k) denotes the inter-carrier interference because of the Doppler frequency and $W(k) = DFT\{w(n)\}$

$$H(k) = \sum_{i=0}^{r-1} h_i e^{j2\pi f_{Di} T} \frac{\sin(\pi f_{Di} T)}{\pi f_{Di} T} e^{-j(2\pi \tau_i/N)k} \tag{8}$$

$$I(k) = \sum_{i=0}^{r-1} \sum_{\substack{K=0 \\ K \ne k}}^{N-1} \frac{h_i X(k)}{N} \frac{1 - e^{j2\pi(f_{Di} T - k + K)}}{1 - e^{j(2\pi/N)(f_{Di} T - k + K)}} e^{-j(2\pi \tau_i/N)K} \tag{9}$$

After passing through the DFT block, the pilot signals are extracted and cross the channel estimation block, then the estimated channel $H_e(k)$ for the data sub-channel is obtained and the transmitted data is estimated by:

$$X_e(k) = \frac{Y(k)}{H_e(k)} \quad k = 0, 1, 2, ..., N-1 \tag{10}$$

Finally, the binary information data is restored back in the signal demapper block.

## 3. CHANNEL ESTIMATION SCHEME

The model adopted for the channel estimation is the block - type pilot insertion in which OFDM channel estimation symbols are transmitted regularly and all sub-carriers are employed as pilots. If the channel is invariable during the block, there will be no error in the channel estimation as the pilots are sent at all carriers. The estimation can be performed by using LS, MMSE or LMMSE algorithms, the Lr-LMMSE algorithm performs well in some cases than the three previous algorithms as will be shown in the simulations.

Assuming that inter symbol interference is dropped by guard interval, we write Y(k) as:

$$Y = Xh + n \qquad (11)$$

Where **y** is the received vector, X is a matrix containing the transmitted signaling points on its diagonal, h is a channel attenuation vector, and n is a vector of i.i.d. complex, zero mean, Gaussian noise with variance $\sigma^2{}_n$

In the following we present the LMMSE estimate of the channel attenuations h from the received vector **y** and the transmitted data X. We assume that the received OFDM symbol contains data known to the estimator - either training data or receiver decisions. [3][9].

The complexity reduction of the LMMSE estimator consists of two separate steps. In the first step we modify the LMMSE by averaging over the transmitted data, obtaining a simplified estimator. In the second step, we reduce the number of multiplications required by applying the theory of optimal rank-reduction [13].

In the third sub section, we will present the Minimum Mean Square Error estimator and we will compare its behaviour with the others algorithms in the section following section.

### 3.1. LMMSE Estimator

The LMMSE estimate of the channel attenuations h, in (11), from the received data **y** and the transmitted symbols X is [14]

$$\hat{h}_{lmmse} = R_{hh_{ls}} R_{h_{ls}h_{ls}}^{-1} \hat{h}_{ls} == R_{hh} \left( R_{hh} + \sigma^2{}_n (XX^h)^{-1} \right)^{-1} \hat{h}_{ls} \qquad (12)$$

Where

$$\hat{h}_{ls} = X^{-1} y = \left[ \frac{y_0}{x_0} \frac{y_1}{x_1} \ldots \frac{y_{N-1}}{x_{N-1}} \right]^T \qquad (13)$$

is the least-squares (LS) estimate of **h**, $\sigma^2{}_n$ is the variance of the additive channel noise, and the covariance matrices are

$$R_{hh} = E\left\{ hh^H \right\} \qquad (14)$$

$$R_{hh_{ls}} = E\left\{ h\, \hat{h}_{ls}^{\;H} \right\} \tag{15}$$

$$R_{h_{ls}h_{ls}} = E\left\{ \hat{h}_{ls}\, \hat{h}_{ls}^{\;H} \right\} \tag{16}$$

The LMMSE estimator (12) is of considerable complexity, since a matrix inversion is needed every time the training data in **X** changes. We reduce the complexity of this estimator by averaging over the transmitted data [10], *i.e.* we replace the term $(XX^h)^{-1}$ in (12) with its expectation $E(XX^h)^{-1}$. Assuming the same signal constellation on all tones and equal probability on all constellation points, we get $E(XX^h)^{-1} = E\left|1/x_k\right|^2 I$, here **I** is the identity matrix. Defining the average signal-to-noise ratio as

$$SNR = E\left|x_k\right|^2 / \sigma^2{}_n \tag{17}$$

, we obtain a simplified estimator

$$\hat{h} = R_{hh}\left( R_{hh} + \frac{\beta}{SNR}I \right)^{-1} \hat{h}_{ls} \tag{18}$$

Where $\beta = E\left|x_k\right|^2 E\left|1/x_k\right|^2$ is a constant depending on the signal constellation. In the case of 16-QAM transmission, $\beta = 17/9$. Because X is no longer a factor in the matrix calculation, no inversion is needed when the transmitted data in X changes. Furthermore, if R$_{hh}$ and SNR are known before hand or are set to fixed nominal values, the matrix $R_{hh}\left( R_{hh} + \dfrac{\beta}{SNR}I \right)^{-1}$ needs to be calculated only once. Under these conditions the estimation requires $N$ multiplications per tone. To further reduce the complexity of the estimator, we proceed with low-rank approximations in the next section.

### 3.2. Optimal Low -rank Approximation

The optimal rank reduction of the estimator in (18), using the singular value decomposition (SVD), is obtained by exclusion of base vectors corresponding to the smallest singular values [13]. We denote the SVD of the channel correlation matrix

$$R_{hh} = D\Lambda D^H \tag{19}$$

Where D is a matrix checking to have orthonormal columns $d_0, d_1, \ldots, d_{N-1}$ and $\Lambda$ designs a diagonal matrix which contains the singular values $\lambda_0 \geq \lambda_1 \geq \ldots \geq \lambda_{N-1} \geq 0$ on its diagonal [13]. This allows the estimator in (18) to be written:

$$\hat{h} = D\Delta D^H \hat{h}_{ls}$$

(20)

Where $\Delta$ is a diagonal matrix containing the values $\delta_k = \dfrac{\lambda_k}{\lambda_k + \dfrac{\beta}{SNR}} k = 0,1...,N-1$

on its diagonal. The best rank -$p$ approximation of the estimator in (18) then becomes

$$\hat{h}_p = U \begin{pmatrix} \Delta_p & 0 \\ 0 & 0 \end{pmatrix} U^H \hat{h}_{ls}$$

(21)

Where $\Delta_p$ is the upper left $p \times p$ corner of $\Delta$.

Viewing the unitary matrix $D^H$ as a transform, the singular value $\lambda_k$ of $R_{hh}$ is the channel energy contained in the $k$th transform coefficient after transforming the LS estimate $\hat{h}_{ls}$. The dimension of the space of essentially time and band-limited signals leads us to the rank needed in the low-rank estimator. In [15] it is shown that this dimension is about $2BT+1$, where $B$ is the one-sided bandwidth and $T$ is the time interval of the signal. Accordingly, the magnitude of the singular values of $\mathbf{R}_{hh}$ should drop rapidly after about $L+1$ large values, where $L$ is the length of the cyclic prefix ($2B = 1 = Ts$, $T = LTs$ and $2BT +1 = L+ 1$).

### 3.3. MMSE Estimator:

For this estimator, we suppose always that the inter symbol interference ISI is dropped by the guard interval, thus the equation of Y given in (7) will be written as:

$$Y = XFh + W$$

(22)

Where

$$X = diag\{X(0), X(1),..., X(N-1)\}$$

$$Y = diag\{Y(0), Y(1),..., Y(N-1)\}^T$$

$$W = diag\{W(0), W(1),..., W(N-1)\}^T$$

$$H = diag\{H(0), H(1),..., H(N-1)\}^T = DFT_N(h)$$

$$F = \begin{pmatrix} W_N^{00} \cdots & W_N^{0(N-1)} \\ \vdots & \ddots & \vdots \\ W_N^{(N-1)0} \cdots & W_N^{(N-1)(N-1)} \end{pmatrix}$$

$$(23)$$

$$W_N^{nk} = \frac{1}{N} e^{-j2\pi(n/N)k}$$

Then, the MMSE estimate of h is given by [10], [12]:

$$(24)$$

$$H_{MMSE} = FR_{hY}R_{YY}^{-1}Y$$

Where

$$(24)$$

$$R_{hY} = E\{hY\} = R_{hh}F^H X^H$$

$$(25)$$

$$R_{YY} = E\{YY\} = XFR_{hh}F^H X^H + \sigma^2 I_N$$

design respectively the cross covariance matrix between h and Y and the auto-covariance matrix of Y. $R_{YY}$ is the auto-covariance matrix of h and $\sigma^2$ is the noise variance $E\{|W(k)|^2\}$

## 4. SIMULATIONS RESULTS

### 4.1. Bit Error Rate

We evaluate the performance of the proposed scheme through the computer simulations and compare it with the conventional schemes. The OFDM system parameters used in our simulation are presented in Table I

| Parameters | Specifications |
|---|---|
| FFT size | 128, 256, 512, 1024,2048 |
| Number of active carriers | 256 |
| Pilot Ratio | 1/8 |
| Guard interval | 256 |
| Guard type | Cyclic extension |
| Bandwidth | 17.5kHz |
| Signal constellation | 16QAM, BPSK |
| Channel Model | Rayleigh fading |

Table I: Simulation Parameters

For this framework, we suppose that we have a perfect synchronization since that our object is to study channel estimation performance.

In order to pass up the inter-symbol interference, the guard interval is chosen to be grater than the maximum delay spread.

Simulations are carried out for different FFT size; different signal to noise (SNR) ratios and for different Doppler spreads.

The channel model used is a Rayleigh fading one and for seeing the effect of fading on block type based and LS,MMSE and LMMSE channel estimation, we have modeled the channel which is time-varying as a block-type pilot based channel estimation scheme.

All blocks contain a fixed number of symbols. Pilots are sent in all sub-carriers of the first symbol of each block and the channel estimation is executed by using either LS, MMSE, LMMSE and Low- rank LMMSE estimations.

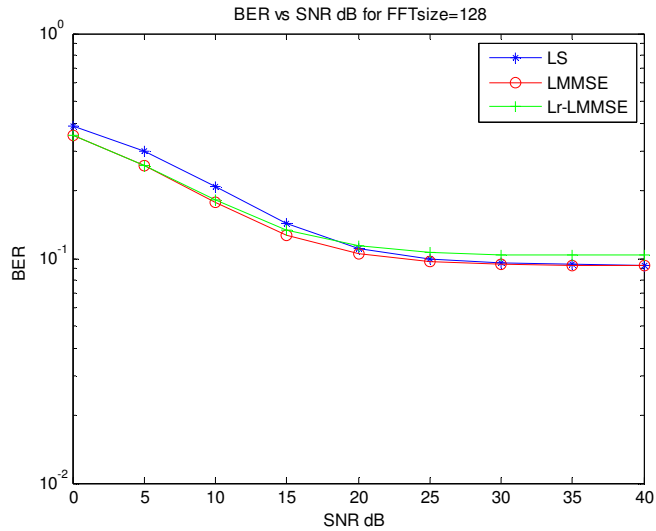

Figure 2: BER Vs SNR for FFT size=128 using LS, LMMSE, Lr-LMMSE algorithms with a 16 QAM modulation

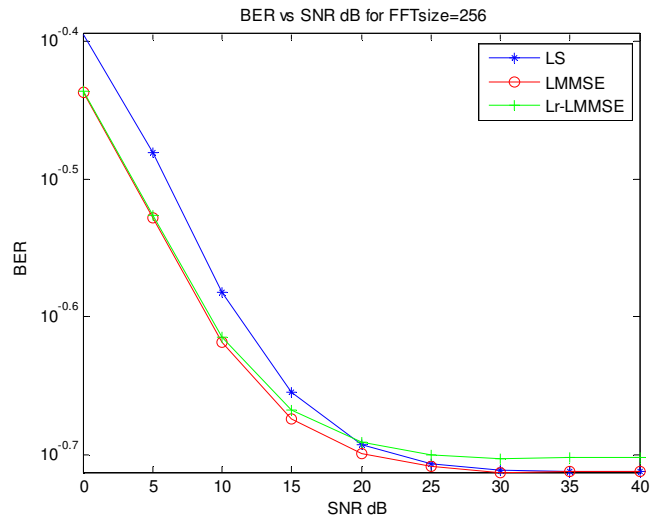

Figure 3: BER Vs SNR for FFT size=256 using LS, LMMSE, Lr-LMMSE algorithms with a 16 QAM modulation

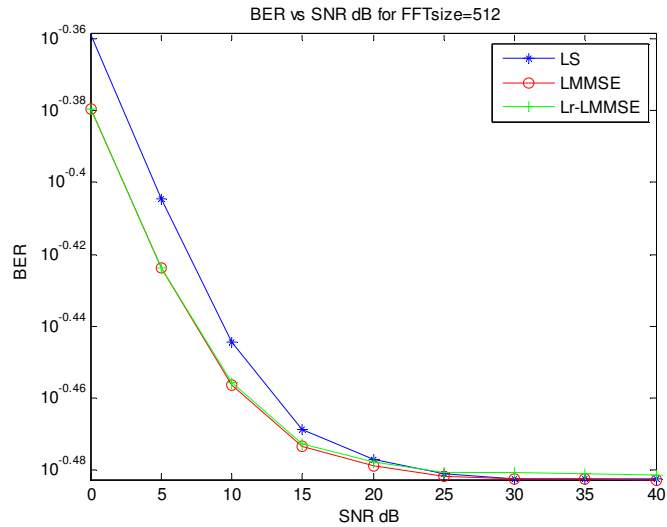

Figure 4: BER Vs SNR for FFT size=512 using LS, LMMSE, Lr-LMMSE algorithms
with a 16 QAM modulation

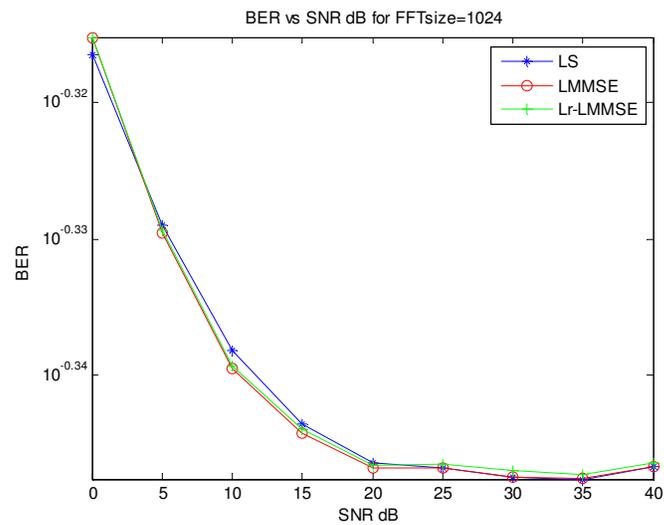

Figure 5: BER Vs SNR for FFT size=1024 using LS, LMMSE, Lr-LMMSE algorithms
with a 16 QAM modulation

Even the FFT size is inferior or equal to 1024, we observe that the LMMSE algorithm still performs well in term of low bit error rate especially when then the SNR is superior to 5 dB. We remark also that by increasing the FFT size, the BER raise for low values of SNR but decreases with values higher than 10 dB.

Finally, we can see that even the LMMSE algorithm performs well with slowly time varying channel i.e with block pilot insertion.

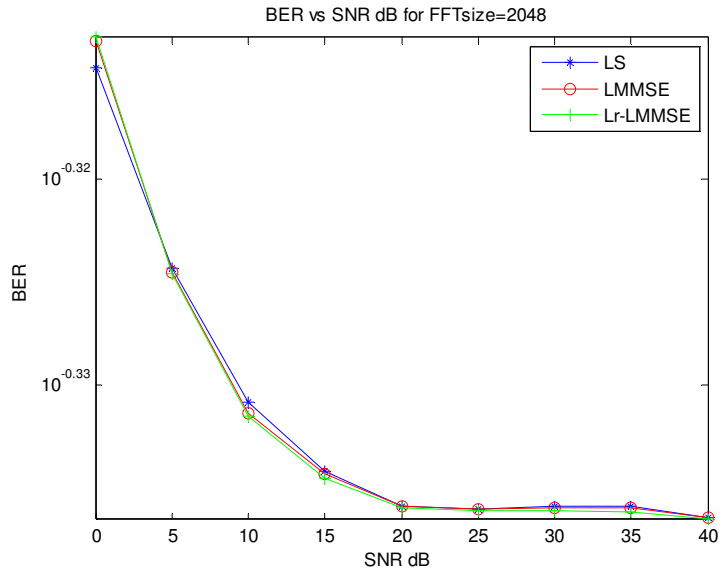

Figure 6: BER Vs SNR for FFT size=2048 using LS, LMMSE, Lr-LMMSE
algorithms with a 16 QAM modulation

When the FFT size is very high, we remark that the three algorithms converge especially when the SNR is higher than 20 dB. For low values of SNR, the three algorithms have a high BER, this is foreseeable because of the effect of the noise. We can see clearly that the Lr-LMMSE estimator performs a few better than the others for an SNR higher than 25 dB.

Now let us check the behaviour of the MMSE algorithm for different sizes of FFT:

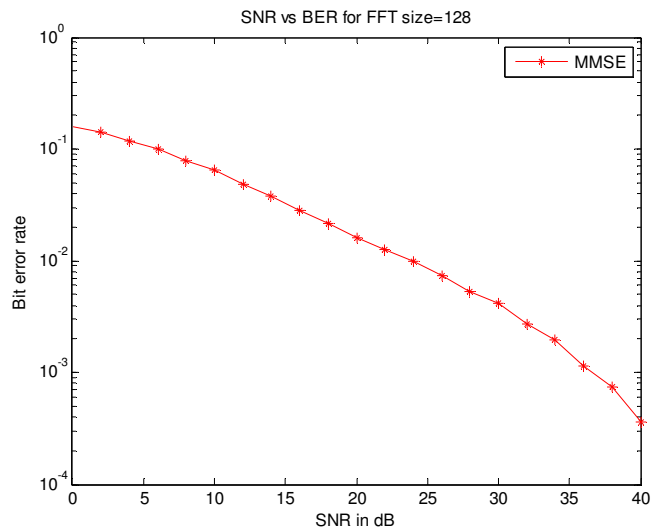

Figure 7: BER vs SNR for FFTsize=128 for the MMSE algorithm with a 16 QAM modulation

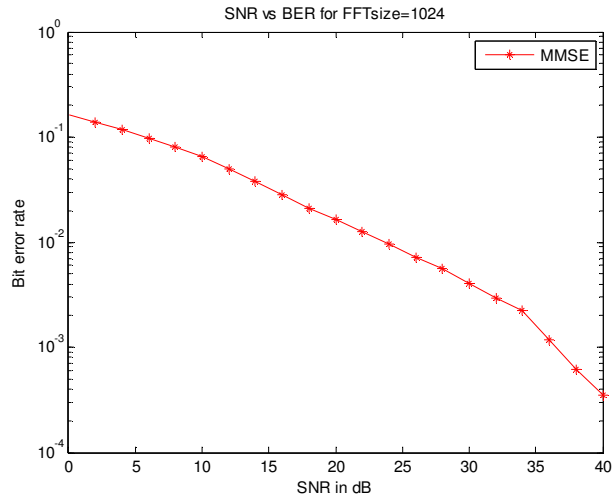

Figure 8: BER vs SNR for FFTsize=1024 for the MMSE algorithm with a 16 QAM modulation

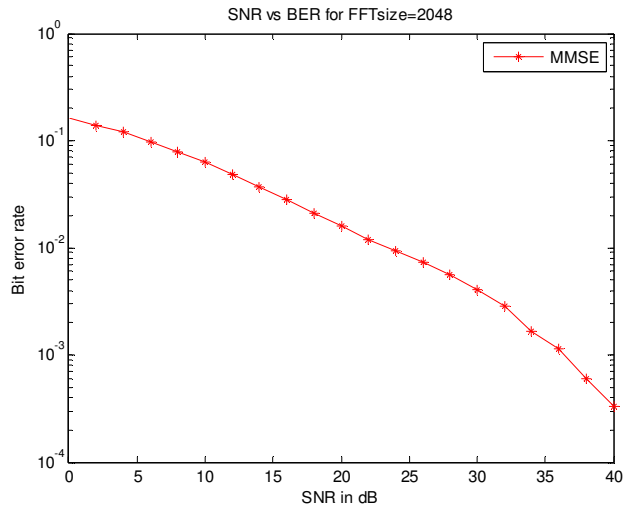

Figure 8: BER vs SNR for FFTsize=2048 with a 16 QAM modulation

From these figures of the BER vs SNR for the MMSE estimator for different sizes of FFT, we observe that this algorithm performs similarly for all sizes of FFT from 128 to 2048. This estimator gives good values for low values of SNR and behaves bad than the others algorithms (LS, LMMSE and Lr-LMMSE) for an SNR higher than 20 dB since they give a BER near to zero for this range of SNR (20dB- 40dB).

Let us now see the representation of the BER vs SNR for a BPSK modulation, this will be shown with the figure behind:

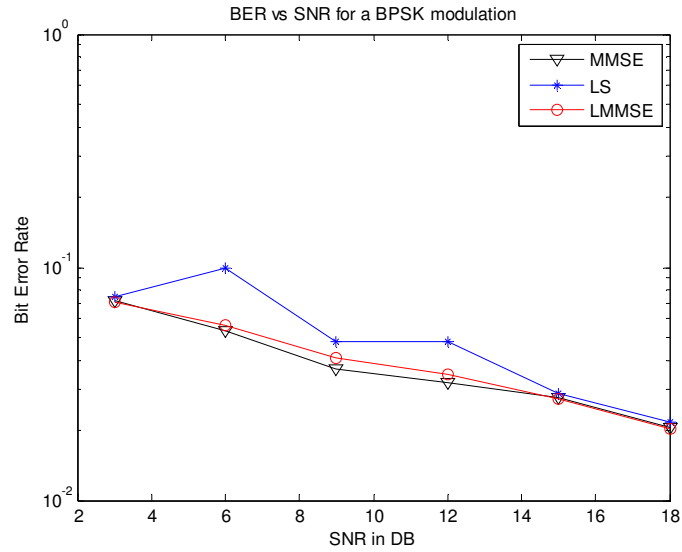

Figure 9: BER vs SNR for a BPSK modulation

The behaviour of the BER for these three algorithms with a BPSK modulation for different sizes of FFT is the same that's why we have represented the main figure for all sizes.

We can mainly observe that for low values of SNR (under 15 dB), the MMSE algorithm performs well but with the top of this value the two estimators (MMSE and LMMSE) behave better than the third one (LS estimator).

## 4.2. Mean Squared Error

The figure below shows the evolution of the mean squared error versus the SNR for the LS, MMSE and LMMSE estimators.

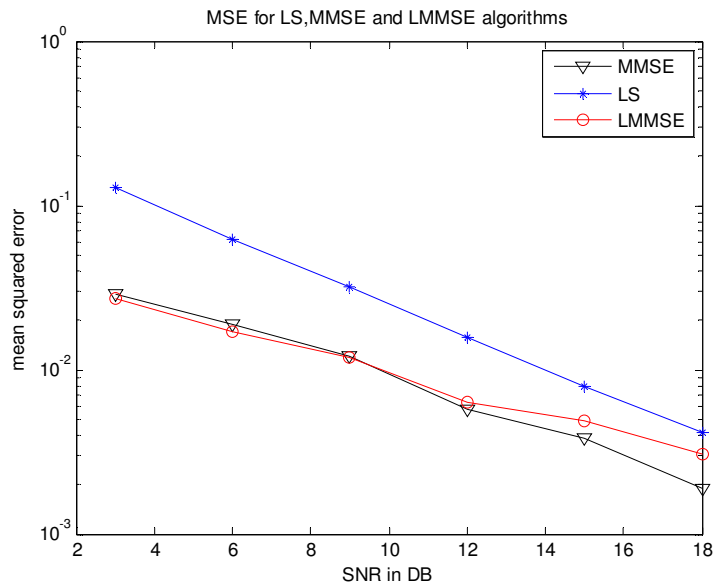

Figure 10: MSE vs SNR for a 16 QAM modulation

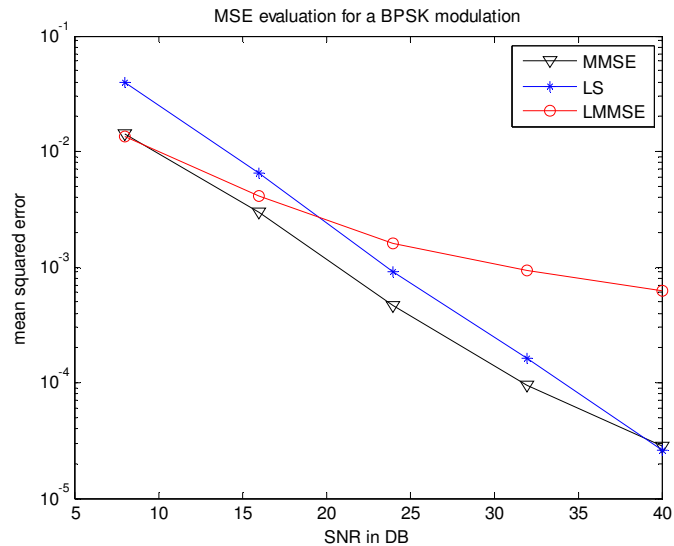

Figure 11: MSE vs SNR for a BPSK modulation

From these figures, we can examine that the behaviour of the LMMSE algorithm with the BPSK modulation is worse than that given by a 16 QAM modulation.

## 5. CONCLUSIONS

In this paper, we proposed a simple and low-complexity approach for the estimation of time varying OFDM channels using the four algorithms LS, MMSE, LMMSE and Lr-LMMSE in the case of block-pilot insertion. This study is an improvement of a our previous work and achieve a hole study of the behaviour of the different estimators proposed with this scheme estimation.
We demonstrated that the LMMSE algorithm is not only convenient to comb – pilot insertion, as mentioned in others publications, but also to block-pilot insertion for the estimation of OFDM channel since it gives a good enhancement of the BER versus SNR and a good MSE.
In fact, it's required to improve this estimator with its low rank version by other schemes of modulations and this will be done in future works.

Nevertheless, let us note that the low rank estimator is shown to be a robust estimator to changes in the channel characteristics and perform very well even when we run it with a high size of the FFT.

I have to invoke the main difficulty behind specially the time of simulations which can attain 48 hours when the FFT size is very high (2048) and also the high computational complexity of MMSE, LMMSE and Lr-LMMSE algorithms.

**Authors**

**Pʀ. RIDHA BOUALLEGUE**
Received the Ph.D degrees in electronic engineering from the National Engineering School of Tunis. In Mars 2003, he received the Hd.R degrees in multiuser detection in wireless communications. From September 1990
He was a graduate Professor in the higher school of communications of Tunis (SUP'COM), he has taught courses in communications and electronics. From 2005 to 2008, he was the Director of the National engineering school of Sousse. In 2006, he was a member of the national committee of science technology. Since 2005, he was the laboratory research in telecommunication Director's at SUP'COM.
From 2005, he served as a member of the scientific committee of validation of thesis and Hd.R in the higher engineering school of Tunis. His recent research interests focus on mobile and wireless communications, OFDM, OFDMA, Long Term Evolution (LTE) Systems. He's interested also in space-time processing for wireless systems and CDMA systems.

**ZAIER AIDA**

Received the B.S. degree in 2005 from National Engineering School of Gabes, Tunisia, and M.S. degree in 2006 from Polytechnic School of Sophia Antipolis of Nice Frrance. Her Research interests focus on channel estimation and synchronization of OFDM and MIMO-OFDM channels under very high mobility conditions.